\documentclass[fleqn,usenatbib]{mnras}

\usepackage{newtxtext,newtxmath}

\usepackage[T1]{fontenc}

\DeclareRobustCommand{\VAN}[3]{#2}
\let\VANthebibliography\thebibliography
\def\thebibliography{\DeclareRobustCommand{\VAN}[3]{##3}\VANthebibliography}

\usepackage{graphicx}	
\usepackage{amsmath}	
\usepackage{natbib}
\usepackage{bm}

\usepackage{color}
\usepackage{ulem}

\def\Vec#1{\mathbfit #1}

\title[Strong toroidal fields in a neutron star]
{Strong toroidal magnetic fields sustained by the elastic crust in a neutron star}
\author[K. Fujisawa et al.]{
Kotaro Fujisawa$^{1}$\thanks{E-mail: fujisawa@resceu.s.u-tokyo.ac.jp},
Yasufumi Kojima$^{2}$ \& 
Shota Kisaka$^{2}$ \\
$^{1}$Department of Physics, Graduate School of Science, the University of Tokyo, Bunkyo-ku, Tokyo 113-0033, Japan,\\
$^{2}$Department of Physics, Graduate School of Advanced Science and Engineering, Hiroshima University, Higashi-Hiroshima, Hiroshima 739-8526, Japan \\
}

\date{Accepted 2022 December 14. Received 2022 December 12; in original form 2022 September 29}

\pubyear{2022}

\begin{document}
\label{firstpage}
\pagerange{\pageref{firstpage}--\pageref{lastpage}}
\maketitle

\begin{abstract}
We investigate new solutions for magnetized neutron stars with a barotropic core in magnetohydrodynamic (MHD) equilibrium and a magneto-elastic crust, which was neglected by previous studies concerning stars in MHD equilibrium. The Lorentz force of the barotropic star is purely irrotational and the structures of magnetic fields are constrained. By contrast, a solenoidal component of the Lorentz force exists in the elastic crust and the structures of the magnetic fields are less restricted. We find that the minor solenoidal component in the elastic crust is important for sustaining the strong magnetic field in the core. Unlike previous studies, the toroidal magnetic field exists in the entire region of the core, and we obtain equilibrium states with large toroidal magnetic fields, where the toroidal magnetic energy is larger than the poloidal magnetic energy. The elastic force of the crust sustains an order of $10^{15}~\mathrm{G}$ toroidal magnetic field in the core, and the maximum strength of the toroidal magnetic field is approximately proportional to the crust thickness. 

\end{abstract}




\section{Introduction}

A neutron star has a strong dipole magnetic field. A magnetar, which is a class of neutron star, has a particularly intense dipole magnetic field that typically reaches approximately $10^{14-15}$ G at its surface. A magnetar is expected to have a more intense toroidal magnetic field inside it. 

Observationally, the presence of the intense toroidal magnetic fields is supported by low-field magnetars (\citealp{Rea_et_al_2010,Rea_et_al_2012}). Although their dipole magnetic fields are much weaker, they display similar burst activities to typical magnetars. Further possible evidence of intense toroidal magnetic fields is the kilo-second Hard X-Ray pulse-phase modulations from magnetars  (\citealp{Makishima_Enoto_et_al_2014, Makishima_et_al_2016, Makishima_et_al_2019, Makishima_et_al_2021a, Makishima_et_al_2021b}). When the toroidal magnetic field is stronger than the poloidal magnetic field, the magnetar becomes prolate and exhibits precession (\citealp{Cutler_2002, Ioka_Sasaki_2004, Haskell_et_al_2008}). If the pulse-phase modulation in Hard X-Ray indicates the precession of the magnetar, then the magnetar would have intense toroidal magnetic fields. 

The intense toroidal magnetic field is also favoured theoretically. According to the linear stability theory, a purely poloidal magnetic field or a purely toroidal magnetic field in the star is unstable (\citealp{Markey_Tayler_1973, Tayler_1973}). Stable magnetic fields should have poloidal and toroidal components (\citealp{Tayler_1980, Akgun_et_al_2013}). Dynamical simulations have revealed the stability of magnetic fields in stars, and twisted-torus magnetic field structures are considered stable configurations (\citealp{Braithwaite_Spruit_2004}). Dynamical simulations suggested the following stability criterion (\citealp{Braithwaite_2009,Duez_Braithwaite_Mathis_2010}; \citealp{Mitchell_et_al_2015})
\begin{align}
    \alpha \frac{{\cal M}_{p} + {\cal M}_{t}}{|W|} \lesssim \frac{{\cal M}_p}{{\cal M}_t + {\cal M}_p} \lesssim 0.8,
    \label{Eq:stability1}
\end{align}
where ${\cal M}_p$, ${\cal M}_t$ and $W$ are the poloidal magnetic field energy, the toroidal magnetic field energy and the gravitational energy, respectively, and $\alpha$ is a dimensionless factor that is of the order of 1000 for neutron stars. Since the ratio of $({\cal M}_p + {\cal M}_t) / |W|$ is an order of $10^{-5}$ even in the case of a magnetar, the criterion becomes
\begin{align}
    0.2 \lesssim \frac{{\cal M}_t}{{\cal M}_t + {\cal M}_p} \lesssim 0.99.
    \label{Eq:stability2}
\end{align}
Stable magnetic field configurations may have intense toroidal components. Similar stability criteria were confirmed by recent dynamical simulations. \citet{2022MNRAS.511.3983S} found that an initially stronger toroidal field setup sustains the magnetic field configurations to the end of the simulations. \citet{Becerra_et_al_2022b} performed a parameter-space study and validated another stability criterion as
\begin{align}
  0.25 \lesssim  \frac{{\cal M}_t}{{\cal M}_p} \lesssim 
  0.5 \sqrt{\left( \frac{\Gamma}{\gamma} - 1\right) \frac{|W|}{{\cal M}_p}}, 
      \label{Eq:stability3}
\end{align}
where $\gamma$ and $\Gamma$ are the polytropic index and the adiabatic index, respectively. The lower bound (but not upper bound) in equation \eqref{Eq:stability3} is the same as in equation \eqref{Eq:stability2}. In all cases, the toroidal component might be stronger than the poloidal component in neutron stars if the magnetic fields are stable.

However, almost all previous studies concerning barotropic magnetohydrodynamic (MHD) equilibrium states have failed to obtain solutions with strong toroidal magnetic fields. Although solutions with twisted-torus configurations have been obtained, typical values of the energy ratios are ${\cal M}_t / ({\cal M}_t + {\cal M}_p) \sim 0.01$ in Newtonian gravity (\citealp{Tomimura_Eriguchi_2005}; \citealp{Yoshida_Eriguchi_2006}; \citealp{Yoshida_Yoshida_Eriguchi_2006}; \citealp{Lander_Jones_2009}; \citealp{Lander_Jones_2012}; \citealp{Fujisawa_Yoshida_Eriguchi_2012}; \citealp{Lander_2013a}; \citealp{Lander_2014}; \citealp{Armaza_et_al_2015}; \citealp{Lander_et_al_2021}), in general relativity with the conformally flat space-time (\citealp{Pili_et_al_2014, Pili_et_al_2015}), and in the fully general relativistic gravity (\citealp{Uryu_et_al_2014, Uryu_et_al_2019}). The toroidal magnetic field of these solutions is confined within a small region inside the star, and the toroidal magnetic field energy is much smaller than the poloidal magnetic field energy. 

By contrast, non-barotropic stellar models have strong toroidal magnetic fields (\citealp{Mastrano_et_al_2011}; \citealp{Mastrano_Melatos_2012}; \citealp{Akgun_et_al_2013}; \citealp{Yoshida_2013}; \citealp{Mastrano_et_al_2013}). The core in neutron stars could be non-barotropic and stably stratified due to chemical composition gradients (\citealt{Reisenegger_2009}). If the neutron star is non-barotropic, the magnetic field configuration is less constrained than that of a barotropic star (\citealt{Reisenegger_2009,Akgun_et_al_2013,Becerra_et_al_2022a,Becerra_et_al_2022b}). \citet{Mastrano_et_al_2011} obtained a stellar model with a strong toroidal magnetic field by using the perturbative method. Although the toroidal magnetic field is confined to a small region, as in barotropic stellar models, the toroidal magnetic field energy can become considerably larger than the poloidal magnetic field energy in non-barotropic stellar models. Thus, non-barotropic stellar models might be a key to realizing stellar models with intense toroidal magnetic fields. 

Some studies have obtained solutions with strong toroidal magnetic fields even for a barotropic star by applying special boundary conditions. \citet{Duez_Mathis_2010} imposed the boundary condition that the magnetic flux vanishes at the stellar surface. Magnetic fields in their models are confined within the stellar interior. They obtained solutions with strong toroidal magnetic fields, which are essentially the same as those of classical models by \citet{Prendergast_1956} and \citet{Woltjer_1959a}. Similar configurations have also been obtained in general relativistic frameworks (\citealp{Ioka_Sasaki_2004}; \citealp{Yoshida_Kiuchi_Shibata_2012}; \citealp{Yoshida_2019}). A surface current is another key to obtaining stellar models with strong toroidal magnetic fields. \citet{Glampedakis_Andersson_Lander_2012} obtained a solution with a toroidal magnetic field by imposing a surface current on the stellar surface as a boundary condition. \citet{Fujisawa_Kisaka_2014} considered an MHD equilibrium core and a Hall equilibrium crust with surface current on the core-crust boundary and obtained solutions with relatively strong toroidal fields. \citet{Ciolfi_Rezzolla_2013} obtained an approximate solution with a strong toroidal magnetic field by neglecting multipole magnetic fields (see also discussion in \citealp{Fujisawa_Eriguchi_2015}; \citealp{Lander_et_al_2021}).

However, these studies concerning non-barotropic stellar models and special boundary conditions did not consider the physical models for the non-barotropic stellar matter or the physical origin of the boundary conditions. The non-barotropic stellar matter results from the chemical-potential gradients in the core, however, the distributions of the chemical potentials were not considered explicitly. The surface current might be related to the crust (\citealp{Fujisawa_Kisaka_2014}), although the elasticity in the crust was neglected. More detailed physical models for a non-barotropic core or the crust are needed to investigate stellar models with strong toroidal magnetic fields. 

The elasticity in the crust is one of the important microphysics component for the magnetic field of the neutron star. The elastic force affects the stability and structures of magnetic fields. \citet{Bera_et_al_2020} performed dynamical simulations and found that the elastic force suppresses MHD instabilities. \citet{Kojima_et_al_2021, Kojima_et_al_2022} investigated the magneto-elastic equilibrium of a neutron star crust and found that the elastic force sustains a strong magnetic field within the crust. A large amount of the magnetic field is associated with the irrotational component of the Lorentz force, which is balanced with the gravity force and pressure gradient. Only a small elastic force is required to balance the solenoidal component of the Lorentz force. As a result, the minor elastic force supports a strong magnetic field in the crust. The distribution of the elastic force in the crust also contributes to the braking torque of the star because it affects the distribution of the currents inside the star and closes the magnetospheric currents (\citealp{Shibata_Kisaka_2021}).

In this paper, we extend our previous models (\citealp{Kojima_et_al_2021, Kojima_et_al_2022}) to the whole star with elastic crust and MHD core. We formulate and calculate new models of magnetized neutron stars with a barotropic MHD core and elastic crust. This paper is organized as follows. In Section 2, we describe the formulation and numerical method. Numerical results are given in Section 3. Section 4 provides discussion and conclusions. 

\section{Formulation and numerical method}

\subsection{Force balance equation}

We consider a magnetized neutron star with an MHD core and elastic crust in equilibrium. We use both spherical $(r,\theta, \varphi)$ and cylindrical $(\varpi,\varphi,z)$ coordinates. The force balance equation of the neutron star in Newtonian gravity is given by
\begin{align}
    -\frac{1}{\rho}\nabla p - \nabla \phi + \frac{1}{\rho}\Vec{f} + \frac{1}{\rho}\Vec{h} = 0,
\end{align}
where $p$, $\rho$, $\phi$ are, the pressure, mass density, and gravitational potential, respectively.  The third term $\Vec{f} = \Vec{j}/c \times \Vec{B}$ is the Lorentz force, and the fourth term $\Vec{h}$ is the elastic force, where $\Vec{j}$ and $\Vec{B}$ are the electric current density and the magnetic field, respectively. 

We assume the magnetic field is stationary and axisymmetry. Then the magnetic field can be described using two scalar functions as 
\begin{align}
  \Vec{B} = \frac{1}{r \sin \theta}  (\nabla \Psi \times \Vec{e}_\varphi) + \frac{S}{r \sin \theta} \Vec{e}_\varphi,
\end{align}
where $\Psi$ is the poloidal magnetic flux function and $S$ is the poloidal current. From Amp\'ere's law, we obtain the following elliptic type equation called the Grad-Shafranov (GS) equation:
\begin{align}
\frac{j_\varphi}{c} = -\frac{1}{4\pi r\sin \theta} \Delta^* \Psi = 
-\frac{1}{4\pi r \sin \theta}
\left[  \frac{\partial^2 \Psi}{\partial r^2} + 
\frac{\sin \theta}{r^2} \frac{\partial}{\partial \theta} \left( 
\frac{1}{\sin \theta} \frac{\partial \Psi}{\partial \theta} \right) \right],
\label{eq:GS}
\end{align}
where $j_\varphi$ is the $\varphi$ component of the electric current density (toroidal current density) and is a source term of this equation. The toroidal current density is constrained by the force-balance equation. We aim to determine the $j_\varphi$ to satisfy the force balance equation.

\subsection{Barotropic MHD equilibrium core}

We assume that the core in the neutron star is barotropic $(p = p (\rho))$ and does not have elastic force  $(\Vec{h} = 0)$ because we treat the core as an ideal fluid. Under this assumption, the $\varphi$ component of the Lorentz force vanishes ($f_\varphi = 0$), and the Lorentz force becomes irrotational $(\nabla \times (\Vec{f}/\rho)) = 0$. The poloidal current $S$ is an arbitrary function of $\Psi$, and the toroidal current density becomes
\begin{align}
     \frac{j_\varphi}{c} = \rho(r) r \sin \theta F'(\Psi) + \frac{S(\Psi)S'(\Psi)}{4\pi r  \sin \theta},
\label{eq:jphi}
\end{align}
where $F(\Psi)$ is another arbitrary function of $\Psi$ and $'$ denotes the derivative with respect to $\Psi$. $F(\Psi)$ is related to the Lorentz force. The force-balance equation is satisfied if the poloidal current and toroidal current density fulfil this constraint \eqref{eq:jphi}. MHD equilibrium state is obtained by fixing the functional forms of $F(\Psi)$ and $S(\Psi)$.

For $F$, we fix the most straightforward form as 
\begin{align}
    F(\Psi) = F_0 \Psi,
    \label{eq:F}
\end{align}
where $F_0$ is a constant. 

We also fix the functional form of $S$. To avoid a discontinuity in the toroidal magnetic field, almost all previous studies have used the following functional form type (e.g., \citealp{Tomimura_Eriguchi_2005}) 
\begin{align}
    S(\Psi) = 
    \begin{cases}
        S_0 (\Psi - \Psi_{\max}) & (\Psi > \Psi_{\max} )\\
        0 & (\Psi \leq \Psi_{\max}),
    \end{cases}
    \label{eq:S1}
\end{align}
where $S_0$ is a constant and $\Psi_{\max}$ is the maximum value of $\Psi$ in the stellar exterior. However, this functional form limits the region where the toroidal magnetic field exists to a small part of the stellar interior. In particular, as the toroidal magnetic field becomes stronger, the area decreases (see figures in \citealp{Lander_Jones_2009}; \citealp{Fujisawa_Eriguchi_2013}; \citealp{Armaza_et_al_2015}). As a result, the toroidal magnetic field energy ${\cal M}_t$ does not become larger than the poloidal magnetic field energy ${\cal M}_p$. Therefore, we use another functional form in this paper as follows:
\begin{align}
    S(\Psi) = S_0 \Psi.
   \label{eq:S2}
\end{align}
This functional form does not limit the region of the toroidal magnetic field to a small part of the core. Using our chosen functional forms in equations~\eqref{eq:F} and \eqref{eq:S2}, the toroidal current density in equation~\eqref{eq:jphi} becomes
\begin{align}
\frac{j_\varphi}{c} = \rho(r) r \sin \theta F_0 + \frac{S_0^2 \Psi}{4\pi r \sin \theta},
\label{eq:j_phi_core}
\end{align}
and the GS equation (equation \ref{eq:jphi}) becomes
\begin{align}
    \Delta^* \Psi + S_0^2 \Psi = -4 \pi F_0 \rho(r) r^2 \sin^2 \theta.
\end{align}
This is a linear equation and we can obtain dipole field solutions where the $\theta$ dependence is $\Psi \propto \sin^2 \theta$ (\citealp{Broderick_Narayan_2008}; \citealp{Duez_Mathis_2010}; \citealp{Fujisawa_Eriguchi_2015}). Throughout this paper, we assume the flux function is purely dipolar ($\Psi(r,\theta) = a(r)\sin^2 \theta$). Then, the poloidal current $S$ and the toroidal current density $j_\varphi$ are respectively separable as $S(r,\theta) = s(r)\sin^2\theta$ and $\frac{j_\varphi}{c} = j(r) \sin \theta$, where $a(r)$, $s(r)$ and $j(r)$ are functions of $r$.

\subsection{Magneto-elastic equilibrium crust}

The crust is limited to the inner crust, where the mass density ranges from $\rho_c = 1.4 \times 10^{14} \mathrm{g~cm^{-3}}$ at the core–crust boundary $r_c$ to the neutron-drip density $\rho_1 = 4 \times 10^{11}\mathrm{g~cm^{-3}}$ at $R$ (\citealt{Kojima_et_al_2022}). We ignore the outer crust and consider the exterior region as the vacuum. 

First, we briefly review the magneto-elastic equilibrium developed in \citet{Kojima_et_al_2021,Kojima_et_al_2022}. The irrotational part of ${\Vec f}$, which is expressed by a gradient of a scalar, may be balanced with a small perturbation of pressure and gravity as in the MHD core. 
In actual, the magnitude associated with the Lorentz force is $B_{0}^2/(4\pi \rho_{c}\Delta r_{cr})$, whereas that with the pressure and gravity is $G_{\rm{N}}\rho_{c}R$, where $\Delta r_{cr}$,$ G_{\rm{N}}$ are, the thickness of the crust, and the gravitational constant.
A ratio is typically $10^{-5}(B/10^{14}{\rm G})^2$, and therefore the Lorentz force may be regarded as a small perturbation to a spherical-stellar structure. Tiny deformations arise but are not explicitly calculated in this paper.
However, in the solenoidal part, a 'curl' of the Lorentz force should be balanced with the elastic force ${\Vec h}$. Thus, we consider a set of approximated equations:
\begin{equation}
({\Vec f}+{\Vec h})_{\varphi}=0, 
\label{eq:elastic_phi}
\end{equation}
\begin{equation}
   [{\nabla}\times \rho^{-1}({\Vec f}+{\Vec h})]_{\varphi}=0.
  \label{eq:rot_elastic}
\end{equation}
We consider the azimuthal component only in equation (\ref{eq:rot_elastic}) since other poloidal components vanish by equation (\ref{eq:elastic_phi}) and the axial symmetry ($\partial_{\varphi}=0$).
 
The {\it i}-th component  $h_{i}$ is expressed by the shear modulus $\mu$ and
elastic displacement $\xi_{i}$ as
\begin{equation}
h_{i}= {\nabla}_{j} \left[\mu 
({\nabla}_{i} \xi ^{j}  +{\nabla}^{j} \xi _{i}) \right],
\end{equation}
where we assume incompressible motion, $\nabla_{i} \cdot \xi^{i}=0$.

Because the elastic force ${\Vec h}$ may participate in the force balance in the neutron star crust, the constraint of the electric current may be relaxed. We can freely determine the distributions of $S$ and $j_\varphi$. We smoothly connect the values of $S$ and $j_\varphi$ at the core-crust boundary such that each distribution becomes $0$ at $R$. To smoothly connect the functions, we use the following quadratic functions of $r$:
\begin{align}
    s(r) = b_0 + b_1 r + b_2 r^2 + b_3 r^3 + b_4 r^4
 \label{eq:sr}
 \end{align}
\begin{align}
     j(r) = c_0 + c_1 r + c_2 r^2 + c_3 r^3 + c_4 r^4,
 \label{eq:jr}
\end{align}
where the coefficients $b_i$ and $c_i$ are determined by connecting $s(r)$, $j(r)$, $d^ns/dr^n$, and $d^nj/dr^n~(n=1,2)$ at the core-crust boundary and the boundary conditions $s = 0$, $j = 0$, $ds/dr=0$, $dj/dr=0$ at $R$. These conditions determine four coefficients in equations \eqref{eq:sr} and \eqref{eq:jr} completely, and the toroidal current density becomes
\begin{align}
\frac{j_\varphi}{c} = j(r) \sin \theta = (c_0 + c_1 r + c_2 r^2 + c_3 r^3 + c_4 r^4) \sin \theta.
\label{eq:j_phi_crust}
\end{align}
Using these quadratic functions, we avoid discontinuities in both poloidal and toroidal magnetic fields at the core-crust boundary and $R$. The force originating from the forms of equations (\ref{eq:sr}) and (\ref{eq:jr}) is balanced with the elastic force $\Vec{h}$ in the elastic crust.

We assume that the shear modulus is approximately proportional to the mass density such that it depends on the radial coordinate, $\mu=\mu(r)$ as 
\begin{equation}
 \mu = \mu_{c} \frac{\rho(r)}{\rho_c},
   \label{eq:mu}
\end{equation}
where $\mu_c =10^{30}~{\rm{erg}}~{\rm{cm}}^{-3}$ is the shear modulus at the core-crust interface.

Since the magnetic fields are described by 
$\Psi =  a(r) \sin^2 \theta$ and $S = s(r) \sin^2 \theta$,
the elastic displacement induced by
the Lorentz force in equations (\ref{eq:elastic_phi}) and
(\ref{eq:rot_elastic}) is expressed by
the Legendre polynomials with $l=2$ only.
We can explicitly write the displacement as
\begin{equation}
\xi_{r}=\frac{6x_{2}}{r^2}P_{2}(\theta),
~~
\xi_{\theta}=\frac{x_{2}^{\prime}}{r}P_{2, \theta}(\theta),
~~
\xi_{\varphi}=rk_{2}P_{2, \theta}(\theta),
\end{equation}
where $x_2(r)$, and  $k_2(r)$ are 
radial functions and this form satisfies the incompressible condition.
Equation (\ref{eq:elastic_phi}) is reduced to the following second-order differential equation:
\begin{equation}
 (\mu r^{4} k_{2}^{\prime})^{\prime} -4\mu r^{2} k_{2}
    =\frac{1}{6\pi}\left( as^{\prime} -a ^{\prime}s\right).    
    \label{klexpd.eqn}    
\end{equation}
Equation (\ref{eq:rot_elastic}) is reduced to a fourth-order differential equation:
\begin{align}
&\left(\frac{(\mu y_{2})^{\prime}}{\rho}\right)^{\prime}
    -\left( \frac{2\mu^{\prime}}{\rho r} +\frac{6\mu}{\rho r^2}
    \right) y_{2}
%
+2\left(\frac{\mu^{\prime}}{\rho r}
    \right)^{\prime}
   \left(x_{2}^{\prime} -\frac{6}{r}x_{2}
    \right)
\nonumber
\\
&=
    \frac{1}{6\pi}
\left[ \frac{4\pi a^{\prime}jr-ss^{\prime}}{r^2 \rho}
-\left(\frac{4\pi ajr-s^2}{r^2 \rho}
 \right)^{\prime} \right] ,
    \label{glexpd.eqn}
\end{align}
where 
\begin{equation}
    x_{2}^{\prime\prime} -\frac{6}{r^2}x_{2}+y_{2}=0 .
\label{flexpd.eqn}
\end{equation}

Since the star is assumed to be spherically symmetric, the boundary conditions for equations (\ref{klexpd.eqn}), (\ref{glexpd.eqn}) and (\ref{flexpd.eqn}) are given by the force balance across the surfaces at $r_c$ and $R$. That is, the shear stress tensors  $\sigma_{ri} ~(i=r,\theta, \varphi)$ vanish since other stresses for the fluid and magnetic field are assumed to be continuous.

The boundary conditions for the radial functions $k_{2}$, 
$x_{2}$, and  $y_{2}$ at $r_{c}$ and $R$ are explicitly written as
\begin{align} 
&  k_{2}^{\prime} =0,
  \label{bcT13}
  \\
  & \left(r^{-2}x_{2} \right)^{\prime}
 = 0,
   \label{bcT11}
 \\
&
2r x_{2} ^{\prime} -12x_{2}  +r^{2} y_{2} =0.
\label{bcT12}
\end{align}

\subsection{Mass density structure of neutron star}

We assume that the neutron star is spherical symmetry, because the deformation due to magnetic fields is considerable small (\citealp{Haskell_et_al_2008}, \citealp{Fujisawa_Kisaka_2014}). We should introduce an equation of state (EOS) to calculate the mass-density distribution of the neutron star. In this paper, we use a polytropic EOS and three realistic EOSs. We use the following analytic relation for the polytropic EOS:
\begin{align}
    p = K\rho^\gamma,
\end{align}
where $p$ and $K$ are the pressure and the polytropic constant, respectively. We set $K = 1.6\times 10^{5}$ and $\gamma = 2$ in cgs units. Then, the maximum mass and the radius become $1.72 M_\odot$ and $1.18 \times 10^{6}$~cm, respectively (\citealp{Shibata_Taniguchi_Uryu_2005, Kiuchi_Yoshida_2008}). 

For the realistic EOSs, we use the zero-temperature EOS tables on the CompOSE website (\citealp{TOK_2015,OHKT_2017}): SKa (\citealp{GR_2015,CNPA_1997,KNPA_1976}) and SLY230a (\citealp{GR_2015,CNPA_1997,DLNP_2009}) as well as SLy EOS (\citealp{Douchin_Haensel_2001}). The maximum masses of the spherical star are $2.22M_\odot$ (SKa), $2.11M_\odot$ (SLY230a), and $2.05M_\odot$ (SLy4), and the radii of a  $1.4M_\odot$ neutron star are $1.29\times 10^6$~cm (SKa), $1.18\times 10^6$~cm (SLY230a), and $1.24 \times 10^6$~cm (SLy4) respectively. A neutron star with SKa EOS or SLy4 EOS has a relatively large radius. We solve the Tolman–Oppenheimer–Volkoff (TOV) equation to obtain the mass density profile $\rho(r)$, although we formulate magnetic fields and elastic force in Newtonian gravity. 

In the numerical computations of the shear modulus $\mu(r)$ in equation (\ref{eq:mu}), we use the approximated mass density $\hat{\rho}$ given by a smooth function (\citealp{Kojima_et_al_2022}) as
\begin{equation}
\frac{\rho(r)}{\rho_c} \sim \hat{\rho}(r) = 
 \left[ 1-\left(1-\left(\frac{\rho_{1}}{\rho_{c}}\right)^{1/\zeta}\right)
 \left( \frac{r-r_{c}}{R-r_{c}}\right)^\eta
\right]^{\zeta},
\end{equation}
where $\zeta$ and $\eta=$1 or 2 are constants that are chosen to fit to
the density profile $\rho(r) / \rho_c$ within the crust ($ r_{c} \le r \le R$). The density profile $\rho(r)/\rho_c$ is well fitted by this smooth function $\hat{\rho}(r)$.

\subsection{Numerical method}

To solve the GS equation, we use Green's function and obtain the integral form (\citealp{Tomimura_Eriguchi_2005}; \citealp{Fujisawa_Kisaka_2014}) as
\begin{align} 
\Psi(r,\theta) &= 2\pi r\sin \theta \sum_{n=1}^{\infty} \frac{P_n^1(\cos \theta)}{n(n+1)} \int_0^{R} f_n(r,\Tilde{r}) \Tilde{r}^2 d\Tilde{r} \nonumber \\
&\times \int_0^{\pi} P_n^1(\cos \Tilde{\theta}) \sin \Tilde{\theta} d\Tilde{\theta} \times \frac{j_\varphi(\Tilde{r}, \Tilde{\theta})}{c},
\label{eq:GS_int}
\end{align}
where $P_n^1$ is the $n$-th associated Legendre function and $f_n(r,\Tilde{r})$ is the following function:
\begin{align}
    f_n(r,\Tilde{r}) = \begin{cases}
        \dfrac{1}{r} \left( \dfrac{\Tilde{r}}{r}\right)^n &  (\Tilde{r}\leq r) \\
        \dfrac{1}{\Tilde{r}} \left( \dfrac{r}{\Tilde{r}} \right)^n & (\Tilde{r} > r).
    \end{cases}
\end{align}
The toroidal current $j_\varphi/c$ is given by equation~\eqref{eq:j_phi_core} within the core ($0 \leq r \leq r_c$) or equation~\eqref{eq:j_phi_crust} within the crust ($r_c \leq r \leq R$). Since the todoridal current density is separable as $\frac{j_\varphi}{c} = j(r) \sin \theta$, equation \eqref{eq:GS_int} is rewritten as
\begin{equation}
\Psi(r,\theta) = \frac{4\pi}{3} r\sin^2 \theta 
\left[
\int_0^{r} \frac{\Tilde{r}^3}{r^2} j(\Tilde{r}) d\Tilde{r}
+ \int_r^{R} r j(\Tilde{r}) d\Tilde{r}
\right].
        \label{eq:GS_int2}
\end{equation}

We use an iterative scheme (\citealp{Fujisawa_Kisaka_2014}) to solve equation \eqref{eq:GS_int2}. First, we set a trial $\Psi$ as an initial guess and calculate $j_\varphi/c$ using equation~\eqref{eq:jphi} (in the core) and equations~\eqref{eq:sr} and \eqref{eq:jr} (in the crust). Next, we solve equation~\eqref{eq:GS_int2} and obtain a new $\Psi$ distribution. We iterate these processes until the value of $\Psi$ converges. To characterize the solutions, we calculate the toroidal and poloidal magnetic energy as
\begin{align}
    {\cal M}_t = \frac{1}{8\pi} \int B_\varphi^2 dV, 
\end{align}
\begin{align}
    {\cal M}_p = \frac{1}{8\pi} \int (B_r^2 + B_\theta^2) dV.
\end{align}
To obtain the displacement $\xi_r$, $\xi_\theta$ and $\xi_\varphi$ in the elastic crust, we numerically solve equations (\ref{klexpd.eqn}), (\ref{glexpd.eqn}) and (\ref{flexpd.eqn}) as boundary value problems by imposing boundary conditions as in equations (\ref{bcT13}), (\ref{bcT11}), and (\ref{bcT12}).

\section{Results}

\subsection{Magnetic field configurations}

\begin{table*}
	\centering
	\caption{Summary of neutron star models and magnetic fields. $\Delta r_{cr}$ is the thickness of the crust, and ${\cal M}_d = \frac{1}{8\pi}B_d^2 \frac{4\pi}{3}R^3$ is a normalized factor of magnetic energy.}
	\label{tab:example_table}
	{\scriptsize
	
	\begin{tabular}{lcccccccc} 
		\hline
		Model & EOS & mass~[$M_\odot$] & $r_c~[10^6~\mathrm{cm}]$ & $R[10^6~\mathrm{cm}]$ & $\Delta r_{cr}/R$ & ${\cal M}_t/({\cal M}_p+{\cal M}_t)$(core) & ${\cal M}_t/({\cal M}_p+{\cal M}_t)$(crust) & $({\cal M}_p + {\cal M}_t) / {\cal M}_d$\\
		\hline
		P0 & polytrope & 1.40 & 1.199 & 1.520 & 2.11E-1 & 0.489 & 0.664 & 5.524\\
		A0 & SKa & 1.40 & 1.169 & 1.244 & 6.05E-2 & 0.540 & 0.634 & 4.740 \\
		B0 & SLY230a & 1.40 & 1.008 & 1.088 & 4.78E-2 & 0.541 & 0.628 & 4.655 \\
		C0 & Sly4 & 1.40 & 1.077 & 1.132 & 4.90E-2 & 0.542 & 0.612 & 4.637 \\
		P1 & polytrope & 1.72 & 1.049 & 1.180 & 1.11E-1 & 0.515 & 0.638 & 4.993 \\
		A1 & SKa & 2.21 & 1.056 & 1.078 & 2.04E-2& 0.550 & 0.621 & 4.527 \\
		B1 & SLY230a & 2.10 & 0.9877 & 1.005 & 1.73E-2 & 0.551 & 0.620 & 4.514 \\
		C1 & Sly4 & 2.05 & 0.9715 & 0.9897  & 1.84E-2 & 0.542 & 0.612 & 4.465 \\
		\hline
	\end{tabular}
	}
	\label{tab:models}
\end{table*}

\begin{figure*}

\includegraphics[width=\columnwidth]{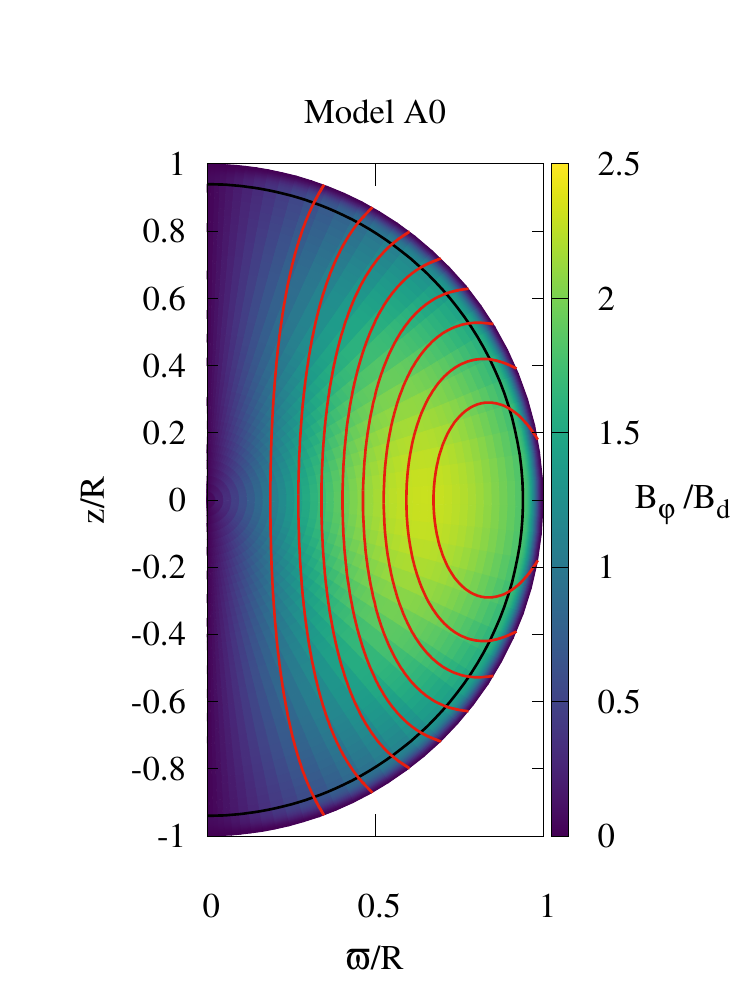}
\includegraphics[width=\columnwidth]{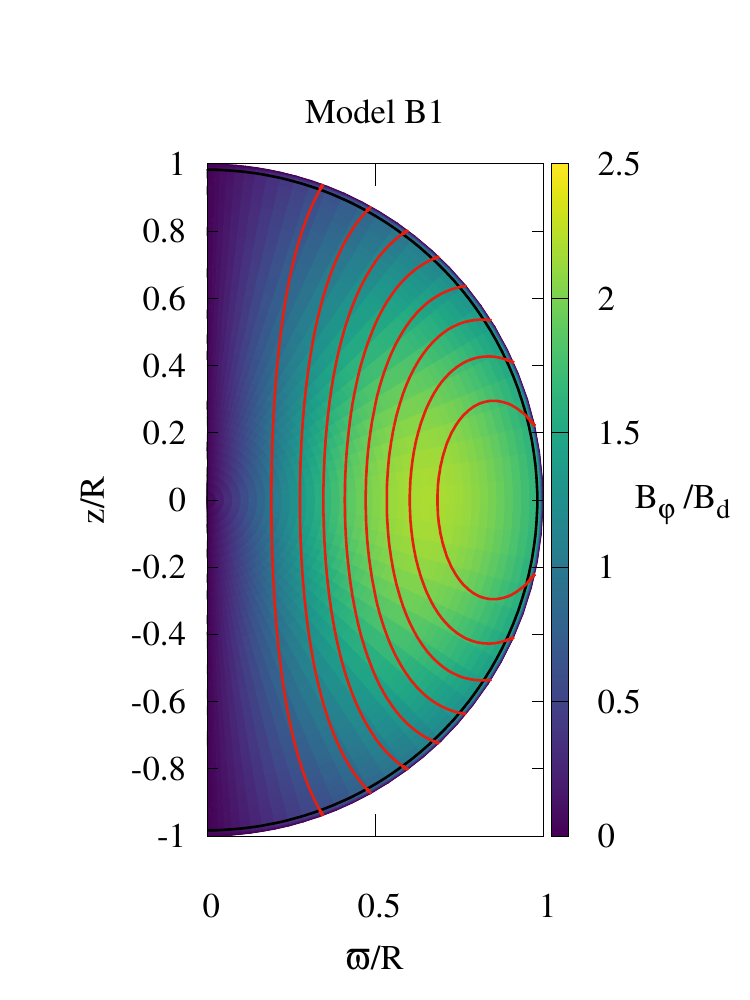}

    \caption{Magnetic field lines and the strength of the toroidal magnetic field (colour maps) in the meridional $\varpi-z$ plane. The strength of the toroidal magnetic field is normalized by the dipole magnetic field at $R$. The inner curve denotes the core-crust boundary.}
    \label{fig:Bphi_map}
\end{figure*}

\begin{figure*}
\includegraphics[width=1.99\columnwidth]{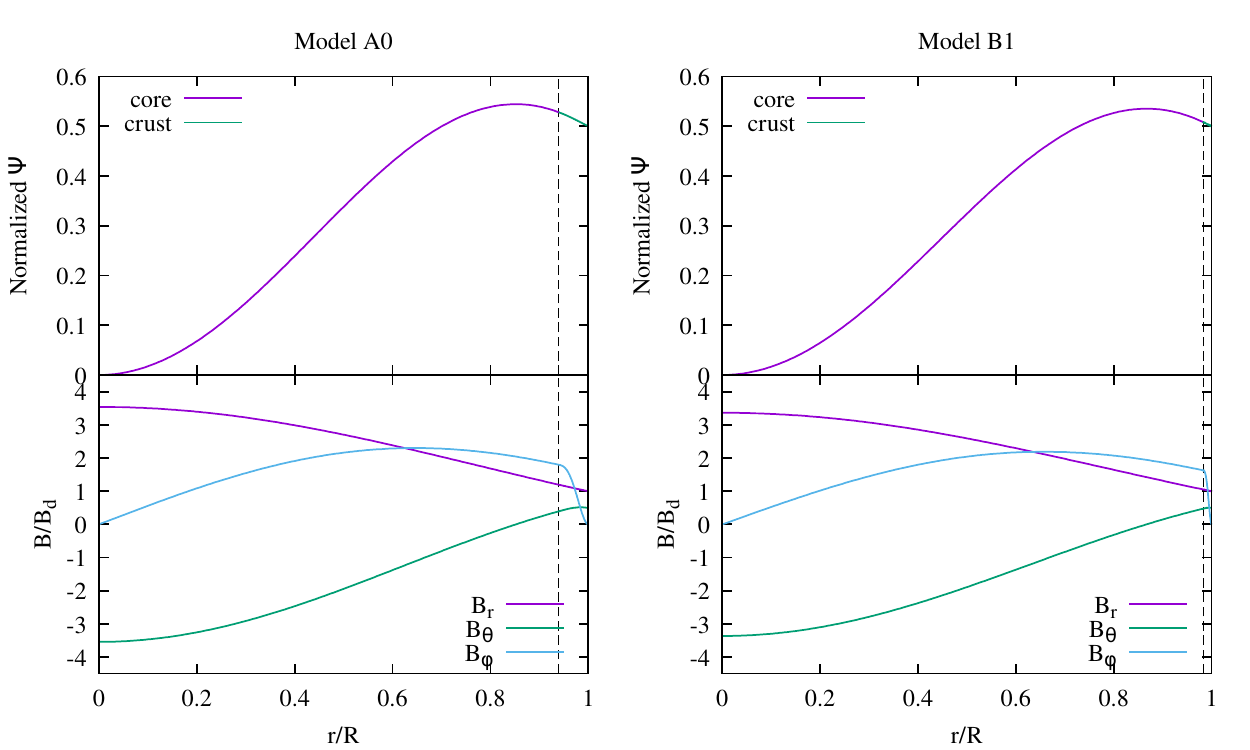}
    \caption{Top: Radial profile of $\Psi$ at $\theta = \pi/2$ of model A0 (left) and model B1 (right). Bottom: Radial profiles of $B_r$ at $\theta = 0$, $B_\theta$ and $B_\varphi$ at $\theta = \pi/2$ for model A0 (left) and model B1 (right). The strength is normalized by $B_d$ (the dipole magnetic field at $R$). The dashed line denotes the location of the core-crust boundary.
    }
    \label{fig:Psi_B}
\end{figure*}

First, we calculate magnetic fields by changing the value of $S_0$ in equation \eqref{eq:S1} and find a critical solution that has the largest value of ${\cal M}_t/({\cal M}_t + {\cal M}_p)$, beyond which the iteration does not converge. The critical solution of each neutron star model is summarized in Table \ref{tab:models}. A model with $0$ indicates a typical neutron star mass ($M = 1.4 M_\odot)$, and a model with $1$ has the maximum mass of each EOS. 

Figure~\ref{fig:Bphi_map} shows the magnetic field lines and the toroidal magnetic fields of model A0 (left) and model B1 (right). The inner curve denotes the core-crust boundary. Although the thickness of the crust is different, the distributions of the magnetic fields are almost the same. The maximum value for the ratio of the toroidal magnetic field to the dipole magnetic field at $R$ is $B_\varphi / B_d \sim 2$, and the region of the toroidal magnetic field is the entire inside the star. This result differs remarkably from previous studies where the toroidal magnetic field is confined within a small part of the star (\citealp{Lander_Jones_2009}; \citealp{Fujisawa_Eriguchi_2013}; \citealp{Armaza_et_al_2015}). In these models, the elastic force in the crust sustains the intense toroidal magnetic field. 

Figure~\ref{fig:Psi_B} displays the radial profiles of the magnetic flux function $\Psi$ (top panels) and magnetic fields (bottom panels) of model A0 and model B1. The dashed line denotes the core-crust boundary. The magnetic flux function is smoothly connected at the core-crust boundary because we choose smooth functional forms for $s$ and $j$ in equations \eqref{eq:sr}. There are no discontinuities in the magnetic fields and these solutions do not have current sheets. 

The toroidal magnetic field energy ${\cal M}_t$ is larger than the poloidal magnetic field energy ${\cal M}_p$ in almost all models. As seen in Table \ref{tab:models}, the values of ${\cal M}_t / ({\cal M}_p + {\cal M}_t)$ are $0.49 - 0.55$ in the core and $0.61 - 0.66$ in the crust. All models satisfy the MHD stability criterion in equation~\eqref{Eq:stability2}. The values of the energy ratio are much higher than those in previous studies (e.g., \citealp{Lander_Jones_2009, Fujisawa_Kisaka_2014}). The thickness of the crust depends on both the EOS and the mass of the neutron star. Except for the polytropic models, model A0 has the thickest crust, whereas model B1 has the thinnest crust. The normalized total magnetic energy $({\cal M}_p + {\cal M}_t) / {\cal M}_d$ is shown in Table \ref{tab:models}, where ${\cal M}_d = \frac{1}{8\pi} B_d^2 \frac{4\pi}{3}R^3$ is a normalized factor of magnetic energy using the dipole magnetic field $B_d$ and $R$. The typical values of the normalized energy are approximately $({\cal M}_p + {\cal M}_t) / {\cal M}_d \sim 5$. The elasticity in the crust sustains relatively large amounts of magnetic energy inside the star. 

\subsection{Irrotational and solenoidal parts of the Lorentz force}

We now check the irrotational and solenoidal components of the Lorentz force in the crust. When the system is barotropic, and the electric current-density is given by equation (\ref{eq:jphi}), the Lorentz force ${\Vec f} \equiv c^{-1}{\Vec j}\times {\Vec B}$ is purely irrotational and expressed by a gradient of a scalar, ${\nabla} F_{0}\Psi$. By contrast, the Lorentz force in the crust has irrotational and solenoidal components and depends on the choice of $s$ and $j$ in equations \eqref{eq:sr} and \eqref{eq:jr}. We consider the effect of our choice of functions in the crust. The Lorentz force is decomposed as a sum of the irrotational and solenoidal parts:
\begin{align}
 \frac{1}{\rho} {\Vec{f}}
 &= -\frac{s}{4\pi \rho r^2} {\nabla}( s \sin^2 \theta)
 + \frac{j}{\rho r}{\nabla}( a \sin^2 \theta)
 \\
 &=
{\nabla} \left( u_{0}(r) + u_{2}(r) P_{2}(\cos \theta) \right)
+
{\nabla} \times \left(  v_{2}(r) \frac{d}{d \theta}P_{2}
 (\cos \theta )\Vec{e}_{\varphi}
 \right),
   \label{eq:iro_sole}
\end{align}
where the radial functions, $u_{0}(r), u_{2}(r)$ and $v_{2}(r)$ are given by
\begin{align}
 & u_{0}(r)  =\frac{2}{3} J_{0}(r),
  \label{eqn.u0}
 \\
& u_{2}(r) =-\frac{4}{5r^3} I_{3}(r)-\frac{4}{15r^3} J_{3}(r)
+\frac{4r^2}{5} I_{-2}(r)-\frac{2r^2}{5} J_{-2}(r),
\label{eqn.u2}
\\
& v_{2}(r)=\frac{2}{5r^3} I_{3}(r)+\frac{2}{15r^3} J_{3}(r)
+\frac{4r^2}{15} I_{-2}(r)-\frac{2r^2}{15} J_{-2}(r).
\label{eqn.v2}
\end{align}
The first and second terms in equation \eqref{eq:iro_sole} are the irrotational and solenoidal parts, respectively. Radial functions $I_{n}$ and $J_{n}$ $(n=-2,0,3)$, which are assumed to vanish at the core-crust boundary $r_{c}$, are given by
\begin{align}
& I_{n}(r) = \int_{r_c} ^{r}
\left( j a  - \frac{{\Tilde{r}}s^{2}}{4\pi}
\right) \frac{{\Tilde{r}}^{n-2}}{\rho} d{\Tilde{r}},
\\
& J_{n}(r) = \int_{r_c} ^{r}
\left( ja^{\prime} - \frac{{\Tilde{r}} s s^{\prime}}{4\pi}
\right) \frac{{\Tilde{r}}^{n-1}}{\rho} d{\Tilde{r}}.
\end{align}

\begin{figure*}

\includegraphics[width=\columnwidth]{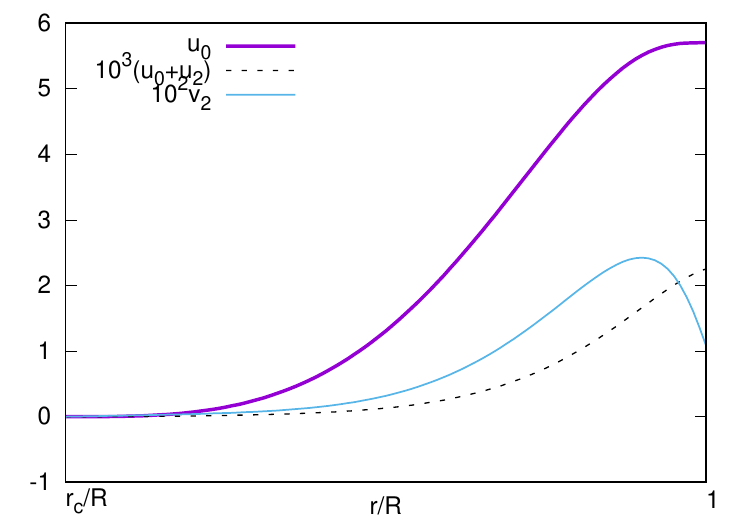}%

\caption{ 
Radial functions $u_{0}, u_{2}$ and $v_{2}$ 
normalized by $B_{0}^2/(4\pi \rho_{c})$ in the crust for Model A0. 
The horizontal axis is a radius normalized by $R$. Function $u_{0}+u_{2}$ is multiplied by $10^3$, and
$v_{2}$ is multiplied by $10^2$ to get the values in a particular scale.
} 
\label{fig:Dem0_pont}

\end{figure*}

Figure \ref{fig:Dem0_pont} shows $u_{0}, u_{2}$ and $v_{2}$ in the crust for model A0 in Table \ref{tab:models}. We see that $u_{0}$ and $u_{2}$ have different signs ($u_0 \sim -u_2$) and that $v_{2}$ relevant to the solenoidal component of the Lorentz force is much smaller in magnitude than that for the irrotational one, $|v_{2}| \ll |u_{0}|, |u_{2}|$. Thus, we may approximate the irrotational component of the Lorentz force as  $\approx {\nabla}(3u_{0} \sin^2 \theta/2) $.
The irrotational component is balanced with the pressure gradient and gravity force in MHD equilibrium because the Lorentz force is much smaller than pressure and gravity.
%
By contrast, the solenoidal component induced by the Lorentz force is still small but not balanced
, because there is no counterpart in the pressure gradient and gravity force
in the barotropic star.
The small solenoidal force is balanced with the elastic force in the crust. Thus, the small elastic force sustains the large magnetic field as in our numerical models.

\subsection{Maximum shear strain}

Finally, we check the maximum shear strain of our models. Using the solution $\xi _{i}$ for equations (\ref{klexpd.eqn}), (\ref{glexpd.eqn}) and (\ref{flexpd.eqn}), we evaluate the shear strain $\sigma_{ij} =({\nabla}_{i} \xi _{j}  +{\nabla}_{j} \xi _{i})/2$. The magnitude $\sigma_{ij}\sigma^{ij}/2$ increases with overall normalization strength $B_{d}$ when the magnetic field configuration is fixed.

\begin{figure*}
\includegraphics[width=2\columnwidth]{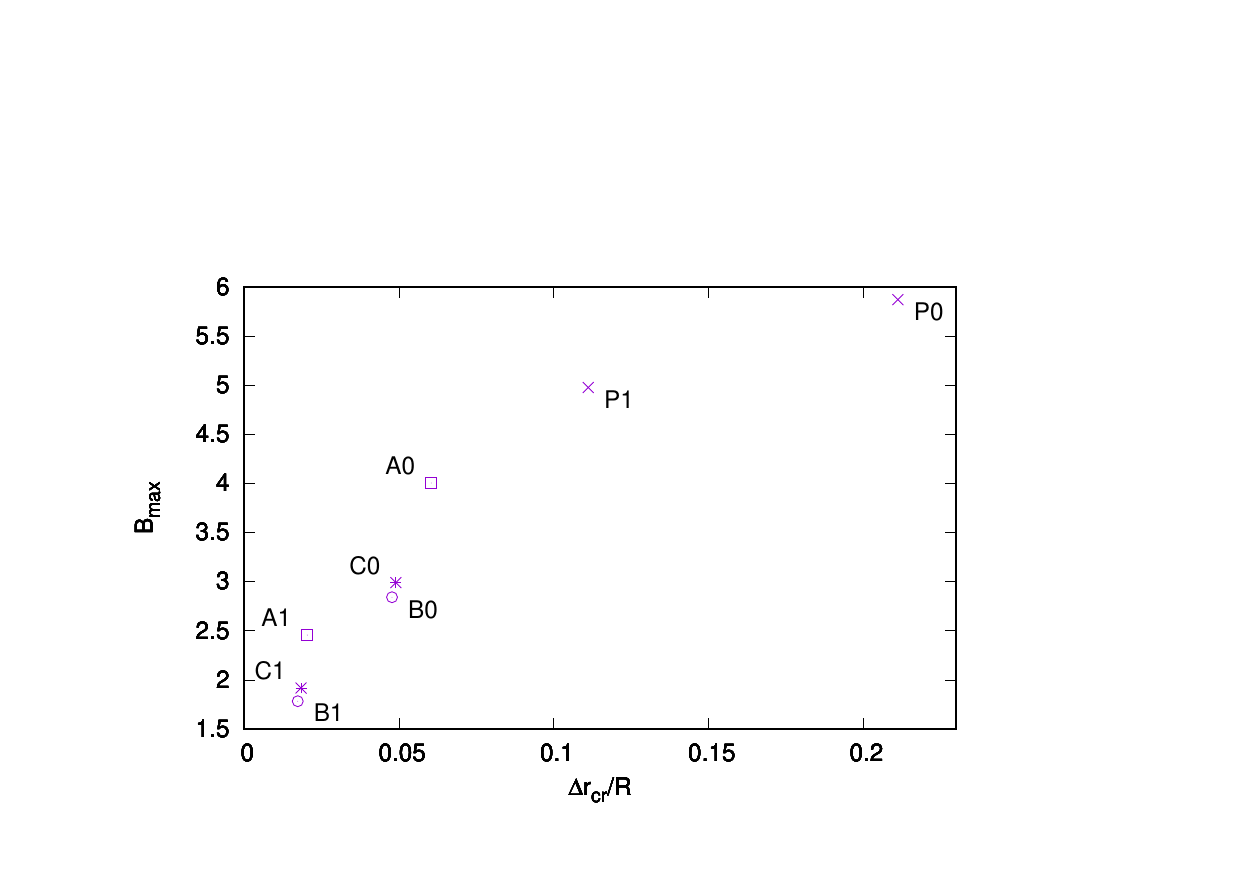}
    \caption{
Maximum magnetic field for the eight models
as a function of crust thickness.
The numerical factor $B_{\rm {max}}$  provides the upper limit of the overall field strength (see equation ~\ref{eq:bmax}).
    }
    \label{fig:maxb14}
\end{figure*}

The crust breaks when the magnitude exceeds a certain threshold. The condition called the Mises criterion is expressed as follows:
\begin{equation}
\frac{1}{2}\sigma_{ij}\sigma^{ij} \le (\sigma_{c})^{2},
   \label{criterion}
\end{equation}
where $\sigma_{c}$ is the maximum strain with a definite value, $\sigma_{c} \approx 10^{-2}-10^{-1}$
(\citealp{2009PhRvL.102s1102H,2018PhRvL.121m2701C,2018MNRAS.480.5511B}). 
This condition provides an upper limit of the overall normalization of the magnetic field $B_{d}$, which is the field strength on the pole,
\begin{equation}
    (B_{d}/10^{14} ~{\rm G}) < B_{\rm {max}}  (\sigma_{c}/0.1)^{1/2},
 \label{eq:bmax} 
\end{equation}
where a numerical factor $B_{\rm {max}}$ depends on the stellar model.
This condition determines the maximum toroidal magnetic field inside the core,
and the maximum magnetic energy sustained by the elastic crust.
The numerical factors $B_{\rm {max}}$ for the eight models in Table \ref{tab:models} are shown in Fig.~\ref{fig:maxb14}. As seen in the figure, 
the maximum field strength generally decreases slightly as the crust thickness decreases. 
Since models with small masses have thick crusts, they have large toroidal magnetic fields in the cores.
Since the values of $B_{\max} \sim 4$ for model A0, the maximum strength of the magnetic fields are $B_d \sim 4\times 10^{14}~\mathrm{G}$ and $B_\varphi \sim 1 \times 10^{15}~\mathrm{G}$.
The maximum total magnetic energy is estimated as 
\begin{equation}
{\cal M}_p + {\cal M}_t = \frac{{\cal M}_p + {\cal M}_t}{{\cal M}_d} \frac{B_d^2}{8\pi} \frac{4\pi}{3}R^3 \sim 2 \times 10^{47} \mathrm{erg}.
\end{equation}

The critical state of the crustal magnetic field was discussed in evolutionary calculations \citep[e.g.,][] {PernPons_2011, Gourgouliatos_Lander_2021}. The crusts in the neutron stars with magnetic fields less than $ \sim 4\times 10^{14}$G failed in their calculations because they used the modified von Mises criterion, in which the elastic deformation was never solved. The approximations result in an incorrect estimate for the elastic limit \citep{Kojima_et_al_2021, Kojima_et_al_2022}.
Moreover, our present limit to the magnetic field strength is based on the force balance in the neutron star, and the magnetic field is close to an MHD equilibrium state. In a secular timescale, however, different modes of elastic deformation far from equilibrium arise during the evolution, even for the weaker magnetic fields, and lead to crustal fracture \citep{Kojima_2022}, although the breakup time significantly increases as the field strength decreases. Magnetized neutron stars may possess the maximum strength given in this paper for some time ($>10^3$ yr).

\begin{figure*}
\includegraphics[width=1.9\columnwidth]{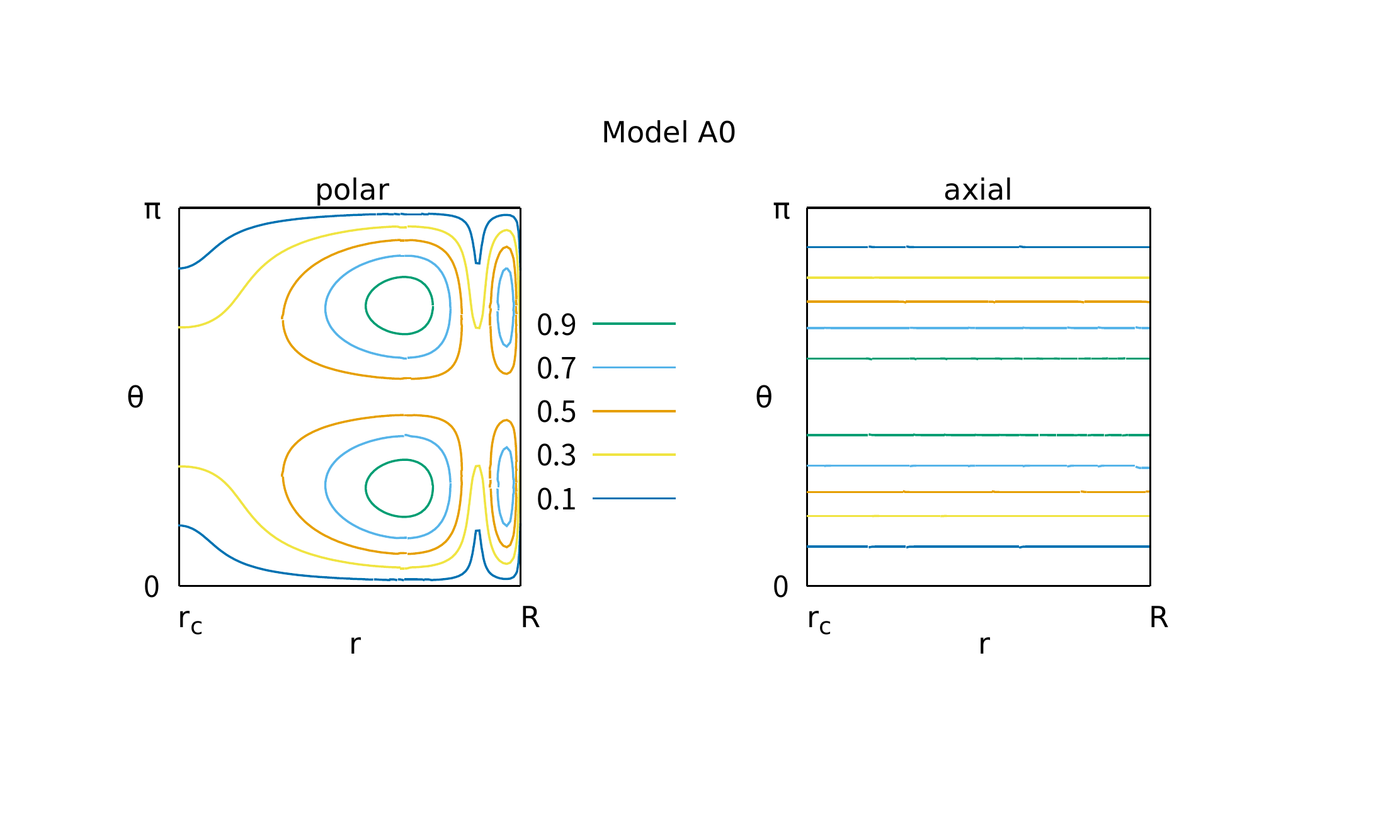}

\includegraphics[width=1.9\columnwidth]{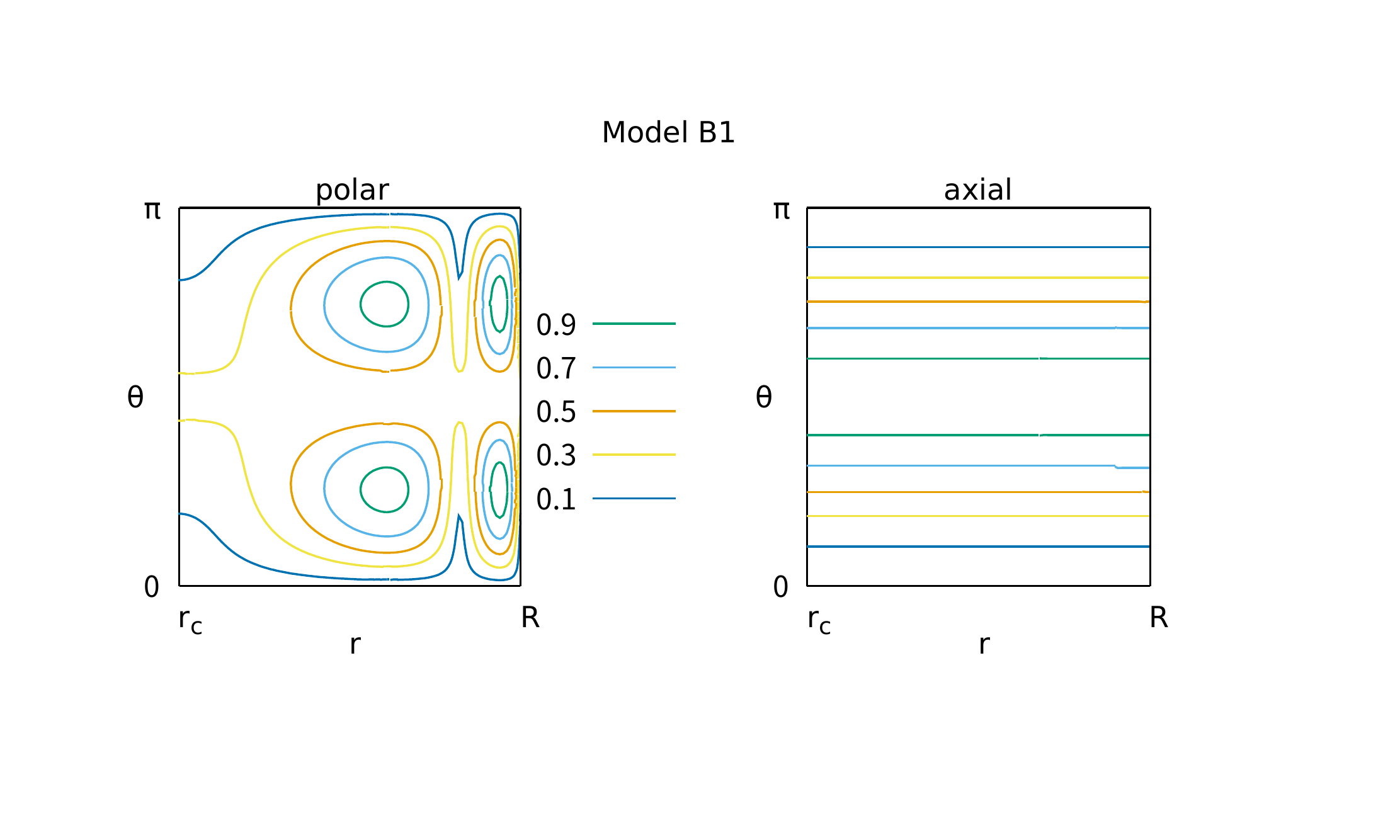}
    \caption{
Magnitude of shear strain normalized by the maximum for model A0 (top) and model B1 (bottom). 
The poloidal part $\sigma_{\rm{pol}}$ and the axial part $\sigma_{\rm{ax}}$ are shown in the crust.} 
    \label{fig:modelA0B1}
\end{figure*}

The shear strain may be split as
\begin{equation}
\frac{1}{2}\sigma_{ij}\sigma^{ij} =  (\sigma_{\rm{pol}})^{2}+ (\sigma_{\rm{ax}})^{2},
%
\end{equation}
where $\sigma_{\rm{pol}}$ denotes a sum of
 $\sigma_{rr}$, $\sigma_{\theta \theta}$, $\sigma_{\varphi \varphi}$
 and $\sigma_{r \theta}$
 induced by the polar displacement, and $\sigma_{\rm{ax}}$ is a sum of $\sigma_{r \varphi}$  and
$\sigma_{\theta \varphi}$ by the axial displacement.
The axial part is larger than the poloidal part; typically, the ratio is
$\sigma_{\rm{pol}} / \sigma_{\rm{ax}} =10^{-5} - 10^{-3}$.
The spatial distribution of $\sigma_{\rm{pol}}$ and $\sigma_{\rm{ax}}$  in the crust
is shown in Fig.\ref{fig:modelA0B1} for Model A0 and Model B1\footnote{
In our previous papers (\citealp{Kojima_et_al_2021, Kojima_et_al_2022}), we used a fixed boundary condition at the core-crust interface and so the spatial profile of the shear stress tensor is different from the present results.
}.
The distribution is quite the similar irrespective of the thickness. 
%
The general features are also similar across the models in Table \ref{tab:models}.
The component $ \sigma_{\theta \phi}$ is the largest, and the angular dependence is simply determined by it with $\sigma_{\theta \phi} \propto \sin^2\theta$.

The dominance of $\sigma_{\rm ax}$ means that $\xi_{\varphi}$ is critical in the maximum magnetic field. Large displacement in the azimuthal direction originates from the large $B_{\varphi}$ component in equation (\ref{klexpd.eqn}).
As the crust becomes thin, $s^\prime$ relevant to the current flow in the crust becomes steep, and the source term increases in magnitude.
Therefore, overall normalization $B_{\rm max}$ for the equilibrium should decrease as shown in Fig.~\ref{fig:maxb14}.

\section{Discussion and conclusions}

We investigated new solutions for magnetized neutron stars with MHD barotropic core and magneto-elastic crust. In the MHD equilibrium core, the Lorentz force is irrotational and the $\varphi$ component of the Lorentz force vanishes. The magnetic fields in the core are characterized by two arbitrary functions $F(\Psi)$ and $I(\Psi)$, in equations~\eqref{eq:F} and \eqref{eq:S1}. In the elastic crust, by contrast, the Lorentz force has a small solenoidal component, and the magnetic field configuration is less constrained. We determined the magnetic field in the crust by using smooth functions in equations~\eqref{eq:sr} and \eqref{eq:jr}, and smoothly connecting all components of the magnetic fields at the boundaries.

In almost all the considered models (Table~\ref{tab:models}), the toroidal magnetic field energy ${\cal M}_t$ is larger than the poloidal magnetic field energy ${\cal M}_p$. Unlike previous studies concerning barotropic stars in MHD equilibrium, the region of the toroidal magnetic field is not limited to a small region inside the star. The toroidal magnetic field in our models exists in the whole region of the core, as seen in Fig.~\ref{fig:Bphi_map}. The minor elastic force of the crust sustains the intense toroidal magnetic field in the core. 

The maximum strength of the toroidal magnetic field is approximately proportional to the crust thickness, which also determines the maximum shear strain of the elastic crust. The relation between the crust thickness and the maximum strength is given in Fig.~\ref{fig:maxb14}. For the case of model A0 with a thick crust, the upper limit of the toroidal magnetic field is in the order of $B_\varphi \sim 1\times 10^{15}~\mathrm{G}$ when the strength of the dipole magnetic field is $B_d \sim 4 \times 10^{14}~\mathrm{G}$. The toroidal magnetic field is sustained by the shear strain of the crust in our models. If the shear strain is increased and exceeds the threshold during the evolution, the toroidal magnetic field can no longer be sustained by the shear strain. Some of the energy of the toroidal magnetic field is released, triggering magnetar-like bursts. As a result, the magnetic fields rearrange and change to a lower energy equilibrium state (\citealp{Pons_Perna_2011}). When the value of the dipole magnetic field is $B_d = 4 \times 10^{14}~\mathrm{G}$, the magnetic energy stored in model A0 is ${\cal M}_p + {\cal M}_t \sim 2 \times 10^{47}~\mathrm{erg}$, some of which is released during the burst event. 

The upper limit of the toroidal magnetic field also depends on magnetic field configurations. We could obtain solutions with a stronger toroidal magnetic field if we choose different functional forms for $F(\Psi)$ and $S(\Psi)$. Further research may extend the models with more intense toroidal magnetic fields. Althoguh the stability and evolution of the magnetic fields of our models are important, they are still unclear. The stability criterion is derived for the fluid star in MHD equilibrium, and the elasticity is neglected. Since the elastic force suppresses the MHD instability (\citealp{Bera_et_al_2020}), our models would be stable. In future works, we aim to construct and systematically analyze the stability. 

\section*{Acknowledgements}

This study was supported by JSPS KAKENHI Grant Number JP20H04728 (KF), JP17H06361, JP19K03850(YK), JP19K14712, JP21H01078, 22H01267, 22K03681 (SK).

\section*{Data Availability}

The data underlying this article will be shared at reasonable request to the corresponding author.



\bibliographystyle{mnras}
\bibliography{ref.bib} 

\begin{thebibliography}{}
\makeatletter
\relax
\def\mn@urlcharsother{\let\do\@makeother \do\$\do\&\do\#\do\^\do\_\do\%\do\~}
\def\mn@doi{\begingroup\mn@urlcharsother \@ifnextchar [ {\mn@doi@}
  {\mn@doi@[]}}
\def\mn@doi@[#1]#2{\def\@tempa{#1}\ifx\@tempa\@empty \href
  {http://dx.doi.org/#2} {doi:#2}\else \href {http://dx.doi.org/#2} {#1}\fi
  \endgroup}
\def\mn@eprint#1#2{\mn@eprint@#1:#2::\@nil}
\def\mn@eprint@arXiv#1{\href {http://arxiv.org/abs/#1} {{\tt arXiv:#1}}}
\def\mn@eprint@dblp#1{\href {http://dblp.uni-trier.de/rec/bibtex/#1.xml}
  {dblp:#1}}
\def\mn@eprint@#1:#2:#3:#4\@nil{\def\@tempa {#1}\def\@tempb {#2}\def\@tempc
  {#3}\ifx \@tempc \@empty \let \@tempc \@tempb \let \@tempb \@tempa \fi \ifx
  \@tempb \@empty \def\@tempb {arXiv}\fi \@ifundefined
  {mn@eprint@\@tempb}{\@tempb:\@tempc}{\expandafter \expandafter \csname
  mn@eprint@\@tempb\endcsname \expandafter{\@tempc}}}

\bibitem[\protect\citeauthoryear{{Akg{\"u}n}, {Reisenegger}, {Mastrano}  \&
  {Marchant}}{{Akg{\"u}n} et~al.}{2013}]{Akgun_et_al_2013}
{Akg{\"u}n} T.,  {Reisenegger} A.,  {Mastrano} A.,   {Marchant} P.,  2013,
  \mn@doi [\mnras] {10.1093/mnras/stt913}, \href
  {http://ads.nao.ac.jp/abs/2013MNRAS.433.2445A} {433, 2445}

\bibitem[\protect\citeauthoryear{{Armaza}, {Reisenegger}  \& {Alejandro
  Valdivia}}{{Armaza} et~al.}{2015}]{Armaza_et_al_2015}
{Armaza} C.,  {Reisenegger} A.,   {Alejandro Valdivia} J.,  2015, \mn@doi
  [\apj] {10.1088/0004-637X/802/2/121}, \href
  {http://adsabs.harvard.edu/abs/2015ApJ...802..121A} {802, 121}

\bibitem[\protect\citeauthoryear{{Baiko} \& {Chugunov}}{{Baiko} \&
  {Chugunov}}{2018}]{2018MNRAS.480.5511B}
{Baiko} D.~A.,  {Chugunov} A.~I.,  2018, \mn@doi [\mnras]
  {10.1093/mnras/sty2259}, \href
  {https://ui.adsabs.harvard.edu/abs/2018MNRAS.480.5511B} {480, 5511}

\bibitem[\protect\citeauthoryear{{Becerra}, {Reisenegger}, {Valdivia}  \&
  {Gusakov}}{{Becerra} et~al.}{2022a}]{Becerra_et_al_2022b}
{Becerra} L.,  {Reisenegger} A.,  {Valdivia} J.~A.,   {Gusakov} M.,  2022a,
  arXiv e-prints, \href {https://ui.adsabs.harvard.edu/abs/2022arXiv220901042B}
  {p. arXiv:2209.01042}

\bibitem[\protect\citeauthoryear{{Becerra}, {Reisenegger}, {Valdivia}  \&
  {Gusakov}}{{Becerra} et~al.}{2022b}]{Becerra_et_al_2022a}
{Becerra} L.,  {Reisenegger} A.,  {Valdivia} J.~A.,   {Gusakov} M.~E.,  2022b,
  \mn@doi [\mnras] {10.1093/mnras/stac102}, \href
  {https://ui.adsabs.harvard.edu/abs/2022MNRAS.511..732B} {511, 732}

\bibitem[\protect\citeauthoryear{{Bera}, {Jones}  \& {Andersson}}{{Bera}
  et~al.}{2020}]{Bera_et_al_2020}
{Bera} P.,  {Jones} D.~I.,   {Andersson} N.,  2020, \mn@doi [\mnras]
  {10.1093/mnras/staa3015}, \href
  {https://ui.adsabs.harvard.edu/abs/2020MNRAS.499.2636B} {499, 2636}

\bibitem[\protect\citeauthoryear{{Braithwaite}}{{Braithwaite}}{2009}]{Braithwaite_2009}
{Braithwaite} J.,  2009, \mn@doi [\mnras] {10.1111/j.1365-2966.2008.14024.x},
  \href {http://} {397, 763}

\bibitem[\protect\citeauthoryear{{Braithwaite} \& {Spruit}}{{Braithwaite} \&
  {Spruit}}{2004}]{Braithwaite_Spruit_2004}
{Braithwaite} J.,  {Spruit} H.~C.,  2004, \mn@doi [\nat] {10.1038/nature02934},
  \href {http://ads.nao.ac.jp/abs/2004Natur.431..819B} {431, 819}

\bibitem[\protect\citeauthoryear{{Broderick} \& {Narayan}}{{Broderick} \&
  {Narayan}}{2008}]{Broderick_Narayan_2008}
{Broderick} A.~E.,  {Narayan} R.,  2008, \mn@doi [\mnras]
  {10.1111/j.1365-2966.2007.12634.x}, \href
  {http://ads.nao.ac.jp/abs/2008MNRAS.383..943B} {383, 943}

\bibitem[\protect\citeauthoryear{{Caplan}, {Schneider}  \& {Horowitz}}{{Caplan}
  et~al.}{2018}]{2018PhRvL.121m2701C}
{Caplan} M.~E.,  {Schneider} A.~S.,   {Horowitz} C.~J.,  2018, \mn@doi [\prl]
  {10.1103/PhysRevLett.121.132701}, \href
  {https://ui.adsabs.harvard.edu/abs/2018PhRvL.121m2701C} {121, 132701}

\bibitem[\protect\citeauthoryear{{Chabanat}, {Bonche}, {Haensel}, {Meyer}  \&
  {Schaeffer}}{{Chabanat} et~al.}{1997}]{CNPA_1997}
{Chabanat} E.,  {Bonche} P.,  {Haensel} P.,  {Meyer} J.,   {Schaeffer} R.,
  1997, \mn@doi [\nphysa] {10.1016/S0375-9474(97)00596-4}, \href
  {https://ui.adsabs.harvard.edu/abs/1997NuPhA.627..710C} {627, 710}

\bibitem[\protect\citeauthoryear{{Ciolfi} \& {Rezzolla}}{{Ciolfi} \&
  {Rezzolla}}{2013}]{Ciolfi_Rezzolla_2013}
{Ciolfi} R.,  {Rezzolla} L.,  2013, \mn@doi [\mnras] {10.1093/mnrasl/slt092},
  \href {http://ads.nao.ac.jp/abs/2013MNRAS.435L..43C} {435, L43}

\bibitem[\protect\citeauthoryear{{Cutler}}{{Cutler}}{2002}]{Cutler_2002}
{Cutler} C.,  2002, \mn@doi [\prd] {10.1103/PhysRevD.66.084025}, \href
  {http://ads.nao.ac.jp/abs/2002PhRvD..66h4025C} {66, 084025}

\bibitem[\protect\citeauthoryear{{Danielewicz} \& {Lee}}{{Danielewicz} \&
  {Lee}}{2009}]{DLNP_2009}
{Danielewicz} P.,  {Lee} J.,  2009, \mn@doi [\nphysa]
  {10.1016/j.nuclphysa.2008.11.007}, \href
  {https://ui.adsabs.harvard.edu/abs/2009NuPhA.818...36D} {818, 36}

\bibitem[\protect\citeauthoryear{{Douchin} \& {Haensel}}{{Douchin} \&
  {Haensel}}{2001}]{Douchin_Haensel_2001}
{Douchin} F.,  {Haensel} P.,  2001, \mn@doi [\aap]
  {10.1051/0004-6361:20011402}, \href
  {http://ads.nao.ac.jp/abs/2001A%26A...380..151D} {380, 151}

\bibitem[\protect\citeauthoryear{{Duez} \& {Mathis}}{{Duez} \&
  {Mathis}}{2010}]{Duez_Mathis_2010}
{Duez} V.,  {Mathis} S.,  2010, \mn@doi [\aap] {10.1051/0004-6361/200913496},
  \href {http://ads.nao.ac.jp/abs/2010A%26A...517A..58D} {517, A58+}

\bibitem[\protect\citeauthoryear{{Duez}, {Braithwaite}  \& {Mathis}}{{Duez}
  et~al.}{2010}]{Duez_Braithwaite_Mathis_2010}
{Duez} V.,  {Braithwaite} J.,   {Mathis} S.,  2010, \mn@doi [\apjl]
  {10.1088/2041-8205/724/1/L34}, \href
  {http://ads.nao.ac.jp/abs/2010ApJ...724L..34D} {724, L34}

\bibitem[\protect\citeauthoryear{{Fujisawa} \& {Eriguchi}}{{Fujisawa} \&
  {Eriguchi}}{2013}]{Fujisawa_Eriguchi_2013}
{Fujisawa} K.,  {Eriguchi} Y.,  2013, \mn@doi [\mnras] {10.1093/mnras/stt541},
  \href {http://ads.nao.ac.jp/abs/2013MNRAS.432.1245F} {432, 1245}

\bibitem[\protect\citeauthoryear{{Fujisawa} \& {Eriguchi}}{{Fujisawa} \&
  {Eriguchi}}{2015}]{Fujisawa_Eriguchi_2015}
{Fujisawa} K.,  {Eriguchi} Y.,  2015, \mn@doi [\pasj] {10.1093/pasj/psv024},
  \href {http://ads.nao.ac.jp/abs/2015PASJ...67...53F} {67, 53}

\bibitem[\protect\citeauthoryear{{Fujisawa} \& {Kisaka}}{{Fujisawa} \&
  {Kisaka}}{2014}]{Fujisawa_Kisaka_2014}
{Fujisawa} K.,  {Kisaka} S.,  2014, \mn@doi [\mnras] {10.1093/mnras/stu1911},
  \href {http://adsabs.harvard.edu/abs/2014MNRAS.445.2777F} {445, 2777}

\bibitem[\protect\citeauthoryear{{Fujisawa}, {Yoshida}  \&
  {Eriguchi}}{{Fujisawa} et~al.}{2012}]{Fujisawa_Yoshida_Eriguchi_2012}
{Fujisawa} K.,  {Yoshida} S.,   {Eriguchi} Y.,  2012, \mn@doi [\mnras]
  {10.1111/j.1365-2966.2012.20614.x}, \href
  {http://ads.nao.ac.jp/abs/2012MNRAS.422..434F} {422, 434}

\bibitem[\protect\citeauthoryear{{Glampedakis}, {Andersson}  \&
  {Lander}}{{Glampedakis} et~al.}{2012}]{Glampedakis_Andersson_Lander_2012}
{Glampedakis} K.,  {Andersson} N.,   {Lander} S.~K.,  2012, \mn@doi [\mnras]
  {10.1111/j.1365-2966.2011.20112.x}, \href
  {http://adsabs.harvard.edu/abs/2012MNRAS.420.1263G} {420, 1263}

\bibitem[\protect\citeauthoryear{{Gourgouliatos} \& {Lander}}{{Gourgouliatos}
  \& {Lander}}{2021}]{Gourgouliatos_Lander_2021}
{Gourgouliatos} K.~N.,  {Lander} S.~K.,  2021, \mn@doi [\mnras]
  {10.1093/mnras/stab1869}, \href
  {https://ui.adsabs.harvard.edu/abs/2021MNRAS.506.3578G} {506, 3578}

\bibitem[\protect\citeauthoryear{{Gulminelli} \& {Raduta}}{{Gulminelli} \&
  {Raduta}}{2015}]{GR_2015}
{Gulminelli} F.,  {Raduta} A.~R.,  2015, \mn@doi [\prc]
  {10.1103/PhysRevC.92.055803}, \href
  {https://ui.adsabs.harvard.edu/abs/2015PhRvC..92e5803G} {92, 055803}

\bibitem[\protect\citeauthoryear{{Haskell}, {Samuelsson}, {Glampedakis}  \&
  {Andersson}}{{Haskell} et~al.}{2008}]{Haskell_et_al_2008}
{Haskell} B.,  {Samuelsson} L.,  {Glampedakis} K.,   {Andersson} N.,  2008,
  \mn@doi [\mnras] {10.1111/j.1365-2966.2008.12861.x}, \href
  {http://adsabs.harvard.edu/abs/2008MNRAS.385..531H} {385, 531}

\bibitem[\protect\citeauthoryear{{Horowitz} \& {Kadau}}{{Horowitz} \&
  {Kadau}}{2009}]{2009PhRvL.102s1102H}
{Horowitz} C.~J.,  {Kadau} K.,  2009, \mn@doi [\prl]
  {10.1103/PhysRevLett.102.191102}, \href
  {https://ui.adsabs.harvard.edu/abs/2009PhRvL.102s1102H} {102, 191102}

\bibitem[\protect\citeauthoryear{{Ioka} \& {Sasaki}}{{Ioka} \&
  {Sasaki}}{2004}]{Ioka_Sasaki_2004}
{Ioka} K.,  {Sasaki} M.,  2004, \mn@doi [\apj] {10.1086/379650}, \href
  {http://ads.nao.ac.jp/abs/2004ApJ...600..296I} {600, 296}

\bibitem[\protect\citeauthoryear{{Kiuchi} \& {Yoshida}}{{Kiuchi} \&
  {Yoshida}}{2008}]{Kiuchi_Yoshida_2008}
{Kiuchi} K.,  {Yoshida} S.,  2008, \mn@doi [\prd] {10.1103/PhysRevD.78.044045},
  \href {http://ads.nao.ac.jp/abs/2008PhRvD..78d4045K} {78, 044045}

\bibitem[\protect\citeauthoryear{{K{\"o}hler}}{{K{\"o}hler}}{1976}]{KNPA_1976}
{K{\"o}hler} H.,  1976, \mn@doi [\nphysa] {10.1016/0375-9474(76)90008-7}, 258,
  301

\bibitem[\protect\citeauthoryear{{Kojima}}{{Kojima}}{2022}]{Kojima_2022}
{Kojima} Y.,  2022, \mn@doi [\apj] {10.3847/1538-4357/ac9184}, \href
  {https://ui.adsabs.harvard.edu/abs/2022ApJ...938...91K} {938, 91}

\bibitem[\protect\citeauthoryear{{Kojima}, {Kisaka}  \& {Fujisawa}}{{Kojima}
  et~al.}{2021}]{Kojima_et_al_2021}
{Kojima} Y.,  {Kisaka} S.,   {Fujisawa} K.,  2021, \mn@doi [\mnras]
  {10.1093/mnras/stab1848}, \href
  {https://ui.adsabs.harvard.edu/abs/2021MNRAS.506.3936K} {506, 3936}

\bibitem[\protect\citeauthoryear{{Kojima}, {Kisaka}  \& {Fujisawa}}{{Kojima}
  et~al.}{2022}]{Kojima_et_al_2022}
{Kojima} Y.,  {Kisaka} S.,   {Fujisawa} K.,  2022, \mn@doi [\mnras]
  {10.1093/mnras/stac036}, \href
  {https://ui.adsabs.harvard.edu/abs/2022MNRAS.511..480K} {511, 480}

\bibitem[\protect\citeauthoryear{{Lander}}{{Lander}}{2013}]{Lander_2013a}
{Lander} S.~K.,  2013, \mn@doi [Physical Review Letters]
  {10.1103/PhysRevLett.110.071101}, \href
  {http://adsabs.harvard.edu/abs/2013PhRvL.110g1101L} {110, 071101}

\bibitem[\protect\citeauthoryear{{Lander}}{{Lander}}{2014}]{Lander_2014}
{Lander} S.~K.,  2014, \mn@doi [\mnras] {10.1093/mnras/stt1894}, \href
  {http://adsabs.harvard.edu/abs/2014MNRAS.437..424L} {437, 424}

\bibitem[\protect\citeauthoryear{{Lander} \& {Jones}}{{Lander} \&
  {Jones}}{2009}]{Lander_Jones_2009}
{Lander} S.~K.,  {Jones} D.~I.,  2009, \mn@doi [\mnras]
  {10.1111/j.1365-2966.2009.14667.x}, \href
  {http://ads.nao.ac.jp/abs/2009MNRAS.395.2162L} {395, 2162}

\bibitem[\protect\citeauthoryear{{Lander} \& {Jones}}{{Lander} \&
  {Jones}}{2012}]{Lander_Jones_2012}
{Lander} S.~K.,  {Jones} D.~I.,  2012, \mn@doi [\mnras]
  {10.1111/j.1365-2966.2012.21213.x}, \href
  {http://ads.nao.ac.jp/abs/2012MNRAS.424..482L} {424, 482}

\bibitem[\protect\citeauthoryear{{Lander}, {Haensel}, {Haskell}, {Zdunik}  \&
  {Fortin}}{{Lander} et~al.}{2021}]{Lander_et_al_2021}
{Lander} S.~K.,  {Haensel} P.,  {Haskell} B.,  {Zdunik} J.~L.,   {Fortin} M.,
  2021, \mn@doi [\mnras] {10.1093/mnras/stab460}, \href
  {https://ui.adsabs.harvard.edu/abs/2021MNRAS.503..875L} {503, 875}

\bibitem[\protect\citeauthoryear{{Makishima}, {Enoto}, {Hiraga}, {Nakano},
  {Nakazawa}, {Sakurai}, {Sasano}  \& {Murakami}}{{Makishima}
  et~al.}{2014}]{Makishima_Enoto_et_al_2014}
{Makishima} K.,  {Enoto} T.,  {Hiraga} J.~S.,  {Nakano} T.,  {Nakazawa} K.,
  {Sakurai} S.,  {Sasano} M.,   {Murakami} H.,  2014, \mn@doi [Physical Review
  Letters] {10.1103/PhysRevLett.112.171102}, \href
  {http://ads.nao.ac.jp/abs/2014PhRvL.112q1102M} {112, 171102}

\bibitem[\protect\citeauthoryear{{Makishima}, {Enoto}, {Murakami}, {Furuta},
  {Nakano}, {Sasano}  \& {Nakazawa}}{{Makishima}
  et~al.}{2016}]{Makishima_et_al_2016}
{Makishima} K.,  {Enoto} T.,  {Murakami} H.,  {Furuta} Y.,  {Nakano} T.,
  {Sasano} M.,   {Nakazawa} K.,  2016, \mn@doi [\pasj] {10.1093/pasj/psv097},
  \href {https://ui.adsabs.harvard.edu/abs/2016PASJ...68S..12M} {68, S12}

\bibitem[\protect\citeauthoryear{{Makishima}, {Murakami}, {Enoto}  \&
  {Nakazawa}}{{Makishima} et~al.}{2019}]{Makishima_et_al_2019}
{Makishima} K.,  {Murakami} H.,  {Enoto} T.,   {Nakazawa} K.,  2019, \mn@doi
  [\pasj] {10.1093/pasj/psy129}, \href
  {https://ui.adsabs.harvard.edu/abs/2019PASJ...71...15M} {71, 15}

\bibitem[\protect\citeauthoryear{{Makishima}, {Enoto}, {Yoneda}  \&
  {Odaka}}{{Makishima} et~al.}{2021a}]{Makishima_et_al_2021a}
{Makishima} K.,  {Enoto} T.,  {Yoneda} H.,   {Odaka} H.,  2021a, \mn@doi
  [\mnras] {10.1093/mnras/stab149}, \href
  {https://ui.adsabs.harvard.edu/abs/2021MNRAS.502.2266M} {502, 2266}

\bibitem[\protect\citeauthoryear{{Makishima}, {Tamba}, {Aizawa}, {Odaka},
  {Yoneda}, {Enoto}  \& {Suzuki}}{{Makishima}
  et~al.}{2021b}]{Makishima_et_al_2021b}
{Makishima} K.,  {Tamba} T.,  {Aizawa} Y.,  {Odaka} H.,  {Yoneda} H.,  {Enoto}
  T.,   {Suzuki} H.,  2021b, \mn@doi [\apj] {10.3847/1538-4357/ac28fd}, \href
  {https://ui.adsabs.harvard.edu/abs/2021ApJ...923...63M} {923, 63}

\bibitem[\protect\citeauthoryear{{Markey} \& {Tayler}}{{Markey} \&
  {Tayler}}{1973}]{Markey_Tayler_1973}
{Markey} P.,  {Tayler} R.~J.,  1973, \mnras, \href
  {http://adsabs.harvard.edu/abs/1973MNRAS.163...77M} {163, 77}

\bibitem[\protect\citeauthoryear{{Mastrano} \& {Melatos}}{{Mastrano} \&
  {Melatos}}{2012}]{Mastrano_Melatos_2012}
{Mastrano} A.,  {Melatos} A.,  2012, \mn@doi [\mnras]
  {10.1111/j.1365-2966.2011.20350.x}, \href
  {http://adsabs.harvard.edu/abs/2012MNRAS.421..760M} {421, 760}

\bibitem[\protect\citeauthoryear{{Mastrano}, {Melatos}, {Reisenegger}  \&
  {Akg{\"u}n}}{{Mastrano} et~al.}{2011}]{Mastrano_et_al_2011}
{Mastrano} A.,  {Melatos} A.,  {Reisenegger} A.,   {Akg{\"u}n} T.,  2011,
  \mn@doi [\mnras] {10.1111/j.1365-2966.2011.19410.x}, \href
  {http://adsabs.harvard.edu/abs/2011MNRAS.417.2288M} {417, 2288}

\bibitem[\protect\citeauthoryear{{Mastrano}, {Lasky}  \& {Melatos}}{{Mastrano}
  et~al.}{2013}]{Mastrano_et_al_2013}
{Mastrano} A.,  {Lasky} P.~D.,   {Melatos} A.,  2013, \mn@doi [\mnras]
  {10.1093/mnras/stt1131}, \href {http://ads.nao.ac.jp/abs/2013MNRAS.434.1658M}
  {434, 1658}

\bibitem[\protect\citeauthoryear{{Mitchell}, {Braithwaite}, {Reisenegger},
  {Spruit}, {Valdivia}  \& {Langer}}{{Mitchell}
  et~al.}{2015}]{Mitchell_et_al_2015}
{Mitchell} J.~P.,  {Braithwaite} J.,  {Reisenegger} A.,  {Spruit} H.,
  {Valdivia} J.~A.,   {Langer} N.,  2015, \mn@doi [\mnras]
  {10.1093/mnras/stu2514}, \href {http://ads.nao.ac.jp/abs/2015MNRAS.447.1213M}
  {447, 1213}

\bibitem[\protect\citeauthoryear{{Oertel}, {Hempel}, {Kl{\"a}hn}  \&
  {Typel}}{{Oertel} et~al.}{2017}]{OHKT_2017}
{Oertel} M.,  {Hempel} M.,  {Kl{\"a}hn} T.,   {Typel} S.,  2017, \mn@doi
  [Reviews of Modern Physics] {10.1103/RevModPhys.89.015007}, \href
  {https://ui.adsabs.harvard.edu/abs/2017RvMP...89a5007O} {89, 015007}

\bibitem[\protect\citeauthoryear{{Perna} \& {Pons}}{{Perna} \&
  {Pons}}{2011}]{PernPons_2011}
{Perna} R.,  {Pons} J.~A.,  2011, \mn@doi [\apjl]
  {10.1088/2041-8205/727/2/L51}, \href
  {http://adsabs.harvard.edu/abs/2011ApJ...727L..51P} {727, L51}

\bibitem[\protect\citeauthoryear{{Pili}, {Bucciantini}  \& {Del Zanna}}{{Pili}
  et~al.}{2014}]{Pili_et_al_2014}
{Pili} A.~G.,  {Bucciantini} N.,   {Del Zanna} L.,  2014, \mn@doi [\mnras]
  {10.1093/mnras/stu215}, \href
  {http://adsabs.harvard.edu/abs/2014MNRAS.439.3541P} {439, 3541}

\bibitem[\protect\citeauthoryear{{Pili}, {Bucciantini}  \& {Del Zanna}}{{Pili}
  et~al.}{2015}]{Pili_et_al_2015}
{Pili} A.~G.,  {Bucciantini} N.,   {Del Zanna} L.,  2015, \mn@doi [\mnras]
  {10.1093/mnras/stu2628}, \href
  {http://adsabs.harvard.edu/abs/2015MNRAS.447.2821P} {447, 2821}

\bibitem[\protect\citeauthoryear{{Pons} \& {Perna}}{{Pons} \&
  {Perna}}{2011}]{Pons_Perna_2011}
{Pons} J.~A.,  {Perna} R.,  2011, \mn@doi [\apj] {10.1088/0004-637X/741/2/123},
  \href {https://ui.adsabs.harvard.edu/abs/2011ApJ...741..123P} {741, 123}

\bibitem[\protect\citeauthoryear{{Prendergast}}{{Prendergast}}{1956}]{Prendergast_1956}
{Prendergast} K.~H.,  1956, \mn@doi [\apj] {10.1086/146186}, \href
  {http://ads.nao.ac.jp/abs/1956ApJ...123..498P} {123, 498}

\bibitem[\protect\citeauthoryear{{Rea} et~al.,}{{Rea}
  et~al.}{2010}]{Rea_et_al_2010}
{Rea} N.,  et~al., 2010, \mn@doi [Science] {10.1126/science.1196088}, \href
  {http://ads.nao.ac.jp/abs/2010Sci...330..944R} {330, 944}

\bibitem[\protect\citeauthoryear{{Rea} et~al.,}{{Rea}
  et~al.}{2012}]{Rea_et_al_2012}
{Rea} N.,  et~al., 2012, \mn@doi [\apj] {10.1088/0004-637X/754/1/27}, \href
  {http://ads.nao.ac.jp/abs/2012ApJ...754...27R} {754, 27}

\bibitem[\protect\citeauthoryear{{Reisenegger}}{{Reisenegger}}{2009}]{Reisenegger_2009}
{Reisenegger} A.,  2009, \mn@doi [\aap] {10.1051/0004-6361/200810895}, \href
  {http://adsabs.harvard.edu/abs/2009A%26A...499..557R} {499, 557}

\bibitem[\protect\citeauthoryear{{Shibata} \& {Kisaka}}{{Shibata} \&
  {Kisaka}}{2021}]{Shibata_Kisaka_2021}
{Shibata} S.,  {Kisaka} S.,  2021, \mn@doi [\mnras] {10.1093/mnras/stab2206},
  \href {https://ui.adsabs.harvard.edu/abs/2021MNRAS.507.1055S} {507, 1055}

\bibitem[\protect\citeauthoryear{{Shibata}, {Taniguchi}  \& {Ury{\=
  u}}}{{Shibata} et~al.}{2005}]{Shibata_Taniguchi_Uryu_2005}
{Shibata} M.,  {Taniguchi} K.,   {Ury{\= u}} K.,  2005, \mn@doi [\prd]
  {10.1103/PhysRevD.71.084021}, \href
  {http://ads.nao.ac.jp/abs/2005PhRvD..71h4021S} {71, 084021}

\bibitem[\protect\citeauthoryear{{Sur}, {Cook}, {Radice}, {Haskell}  \&
  {Bernuzzi}}{{Sur} et~al.}{2022}]{2022MNRAS.511.3983S}
{Sur} A.,  {Cook} W.,  {Radice} D.,  {Haskell} B.,   {Bernuzzi} S.,  2022,
  \mn@doi [\mnras] {10.1093/mnras/stac353}, \href
  {https://ui.adsabs.harvard.edu/abs/2022MNRAS.511.3983S} {511, 3983}

\bibitem[\protect\citeauthoryear{{Tayler}}{{Tayler}}{1973}]{Tayler_1973}
{Tayler} R.~J.,  1973, \mnras, \href
  {http://adsabs.harvard.edu/abs/1973MNRAS.161..365T} {161, 365}

\bibitem[\protect\citeauthoryear{{Tayler}}{{Tayler}}{1980}]{Tayler_1980}
{Tayler} R.~J.,  1980, \mnras, \href
  {http://adsabs.harvard.edu/abs/1973MNRAS.161..365T} {191, 151}

\bibitem[\protect\citeauthoryear{{Tomimura} \& {Eriguchi}}{{Tomimura} \&
  {Eriguchi}}{2005}]{Tomimura_Eriguchi_2005}
{Tomimura} Y.,  {Eriguchi} Y.,  2005, \mn@doi [\mnras]
  {10.1111/j.1365-2966.2005.08967.x}, \href
  {http://ads.nao.ac.jp/abs/2005MNRAS.359.1117T} {359, 1117}

\bibitem[\protect\citeauthoryear{{Typel}, {Oertel}  \& {Kl{\"a}hn}}{{Typel}
  et~al.}{2015}]{TOK_2015}
{Typel} S.,  {Oertel} M.,   {Kl{\"a}hn} T.,  2015, \mn@doi [Physics of
  Particles and Nuclei] {10.1134/S1063779615040061}, \href
  {https://ui.adsabs.harvard.edu/abs/2015PPN....46..633T} {46, 633}

\bibitem[\protect\citeauthoryear{{Ury\=u}, {Gourgoulhon}, {Markakis},
  {Fujisawa}, {Tsokaros}  \& {Eriguchi}}{{Ury\=u}
  et~al.}{2014}]{Uryu_et_al_2014}
{Ury\=u} K.,  {Gourgoulhon} E.,  {Markakis} C.~M.,  {Fujisawa} K.,  {Tsokaros}
  A.,   {Eriguchi} Y.,  2014, \mn@doi [\prd] {10.1103/PhysRevD.90.101501},
  \href {http://adsabs.harvard.edu/abs/2014PhRvD..90j1501U} {90, 101501}

\bibitem[\protect\citeauthoryear{{Ury{\={u}}}, {Yoshida}, {Gourgoulhon},
  {Markakis}, {Fujisawa}, {Tsokaros}, {Taniguchi}  \& {Eriguchi}}{{Ury{\={u}}}
  et~al.}{2019}]{Uryu_et_al_2019}
{Ury{\={u}}} K.,  {Yoshida} S.,  {Gourgoulhon} E.,  {Markakis} C.,  {Fujisawa}
  K.,  {Tsokaros} A.,  {Taniguchi} K.,   {Eriguchi} Y.,  2019, \mn@doi [\prd]
  {10.1103/PhysRevD.100.123019}, \href
  {https://ui.adsabs.harvard.edu/abs/2019PhRvD.100l3019U} {100, 123019}

\bibitem[\protect\citeauthoryear{{Woltjer}}{{Woltjer}}{1959}]{Woltjer_1959a}
{Woltjer} L.,  1959, \mn@doi [\apj] {10.1086/146731}, \href
  {http://ads.nao.ac.jp/abs/1959ApJ...130..400W} {130, 400}

\bibitem[\protect\citeauthoryear{{Yoshida}}{{Yoshida}}{2013}]{Yoshida_2013}
{Yoshida} S.,  2013, \mn@doi [\mnras] {10.1093/mnras/stt1362}, \href
  {http://adsabs.harvard.edu/abs/2013MNRAS.435..893Y} {435, 893}

\bibitem[\protect\citeauthoryear{{Yoshida}}{{Yoshida}}{2019}]{Yoshida_2019}
{Yoshida} S.,  2019, \mn@doi [\prd] {10.1103/PhysRevD.99.084034}, \href
  {https://ui.adsabs.harvard.edu/abs/2019PhRvD..99h4034Y} {99, 084034}

\bibitem[\protect\citeauthoryear{{Yoshida} \& {Eriguchi}}{{Yoshida} \&
  {Eriguchi}}{2006}]{Yoshida_Eriguchi_2006}
{Yoshida} S.,  {Eriguchi} Y.,  2006, \mn@doi [\apj] {10.1086/501050}, \href
  {http://ads.nao.ac.jp/abs/2006ApJS..164..156Y} {164, 156}

\bibitem[\protect\citeauthoryear{{Yoshida}, {Yoshida}  \& {Eriguchi}}{{Yoshida}
  et~al.}{2006}]{Yoshida_Yoshida_Eriguchi_2006}
{Yoshida} S.,  {Yoshida} S.,   {Eriguchi} Y.,  2006, \mn@doi [\apj]
  {10.1086/507513}, \href {http://ads.nao.ac.jp/abs/2006ApJ...651..462Y} {651,
  462}

\bibitem[\protect\citeauthoryear{{Yoshida}, {Kiuchi}  \& {Shibata}}{{Yoshida}
  et~al.}{2012}]{Yoshida_Kiuchi_Shibata_2012}
{Yoshida} S.,  {Kiuchi} K.,   {Shibata} M.,  2012, \mn@doi [\prd]
  {10.1103/PhysRevD.86.044012}, \href
  {http://ads.nao.ac.jp/abs/2012PhRvD..86d4012Y} {86, 044012}

\makeatother
\end{thebibliography}






\bsp	
\label{lastpage}
\end{document}